\documentclass[prd,twocolumn,nopacs,floatfix,amsmath,nofootinbib,amssymb,preprintnumbers]{revtex4-2}
\usepackage{graphicx,subfigure,color,dcolumn,booktabs,diagbox}
\usepackage{txfonts}
\usepackage{slashed}
\usepackage{epstopdf}
\usepackage{multirow}
\usepackage{bm}
\usepackage{pifont}
\usepackage{adjustbox}
\usepackage{siunitx}
\usepackage{multirow}
\usepackage{makecell}
\usepackage{rotating}
\usepackage{verbatim}
\usepackage{graphicx,xcolor,dcolumn,booktabs}
\usepackage[colorlinks,
            citecolor=blue,
            anchorcolor=red,
            menucolor=red,
            linkcolor=red,
            filecolor=red,
            runcolor=red,
            urlcolor=blue,
            frenchlinks=true]{hyperref}
\begin{document}
\author{Zi-Long Man$^{1,2,3,4}$}
\author{Si-Qiang Luo$^{1,2,3,4}$}
\author{Xiang Liu$^{1,2,3,4}$}
\email{xiangliu@lzu.edu.cn}

\affiliation{
$^1$School of Physical Science and Technology, Lanzhou University, Lanzhou 730000, China\\
$^2$Lanzhou Center for Theoretical Physics,
Key Laboratory of Theoretical Physics of Gansu Province,
Key Laboratory of Quantum Theory and Applications of MoE,
Gansu Provincial Research Center for Basic Disciplines of Quantum Physics, Lanzhou University, Lanzhou 730000, China\\
$^3$MoE Frontiers Science Center for Rare Isotopes, Lanzhou University, Lanzhou 730000, China\\
$^4$Research Center for Hadron and CSR Physics, Lanzhou University and Institute of Modern Physics of CAS, Lanzhou 730000, China}
\title{Is the $3S$-$2D$ mixing strong for the charmonia $\psi(4040)$ and $\psi(4160)$?
}

\begin{abstract}
In this work, we revisit the $3S$-$2D$ mixing scheme for the charmonia $\psi(4040)$ and $\psi(4160)$. We introduce a coupled-channel mechanism-distinct from the tensor-force contribution in potential models, which alone is insufficient to induce significant mixing-to describe the mixing between these states. Our analysis yields  mixing angles of $\theta_1=7^\circ$ and $\theta_2=10^\circ$, inconsistent with the larger angle inferred from experimental data, such as the dilectronic widths of the $\psi(4040)$ and $\psi(4160)$. We discuss possible origins of this discrepancy and emphasize the need for future experiments to resolve it. Precise measurements of the resonance parameters and dilectronic  decay widths, via both inclusive and exclusive processes, will be crucial in clarifying this issue.
\end{abstract}

\maketitle

\section{Introduction}

Although 50 years have passed since the discovery of the $J/\psi$, the properties of charmonium states above the $D\bar{D}$ threshold remain poorly understood, presenting a more complex picture due to insufficient experimental data and a lack of comprehensive theoretical understanding. Among these higher-mass states, the $\psi(4040)$ and $\psi(4160)$ were first observed in $e^+e^-$ annihilation by the DASP Collaboration in 1978 \cite{DASP:1978dns}. Both states are now well-established charmonium resonances and are listed in the Review of Particle Physics by the Particle Data Group (PDG). This inclusion might suggest that their properties are sufficiently understood-yet in reality, the situation remains unclear, as we will discuss below.

$S$-$D$ mixing is a universal phenomenon in charmonium physics. A classic example is the $2S$-$1D$ mixing scheme proposed by Rosner for $\psi(3770)$ and $\psi(3686)$ \cite{Rosner:2001nm}, which successfully explains the relevant experimental data. Recently, the Lanzhou group suggested $4S$-$3D$ mixing as a mechanism to interpret the charmoniumlike states $Y(4220)$ and $Y(4380)$ \cite{Man:2025zfu}. Similarly, for the well-established charmonia $\psi(4040)$ and $\psi(4160)$, a $3S$-$2D$ mixing scheme has also been proposed, as discussed in Refs. \cite{Chao:2007it,Li:2009zu, Badalian:2008dv,Anwar:2016mxo,Yang:2018mkn,Wang:2022jxj,Bokade:2024tge,Zhao:2023hxc}. 
The mixing angles can be roughly obtained by fitting either the measured decay widths of the relevant states or the ratio between them.
In Table \ref{mix-angle}, we summarize the current understanding of $3S$-$2D$ mixing, including the inferred mixing angles and the corresponding approaches employed to extract them.
\begin{table}[!htbp]
\caption{The obtained $3S$-$2D$ mixing angle for the $\psi(4040)$ and $\psi(4160)$.}\centering\label{mix-angle}
\renewcommand\arraystretch{1.35}
\begin{tabular*}{1.0\columnwidth}{@{\extracolsep{\fill}}ccr}
\toprule[1.00pt]
\toprule[1.00pt]
$\Gamma_{e^+e^-}^{\psi(4040)}$~(keV) & $\Gamma_{e^+e^-}^{\psi(4160)}$ (keV)  &\multicolumn{1}{c}{Mixing angle}                                             \\
\midrule[0.75pt]
\multirow{6}{*}{$0.86\pm0.07$~\cite{ParticleDataGroup:2024cfk}}     &\multirow{3}{*}{$0.83\pm0.08$~\cite{Seth:2004py}}         &  $\phi=-35^{\circ},+55^{\circ}$~\cite{Chao:2007it} \footnotemark[1]        \\
                                    &                                       &  
                                    $\phi=-37^{\circ}$~\cite{Li:2009zu} \footnotemark[1] \\
                                    &                                       &      $\theta=34.8^{\circ},-55.7^{\circ}$~\cite{Badalian:2008dv} \footnotemark[1]                  \\
\Xcline{2-3}{0.75pt}
                                    &\multirow{3}{*}{$0.48\pm0.22$~\cite{BES:2007zwq}}         &  $\theta=20^{\circ}$~\cite{Wang:2022jxj} \footnotemark[2]                    \\
                                    &                                       &  $\phi=45^{\circ}$~\cite{Bokade:2024tge} \footnotemark[3]                  \\
                                    &                                       &  $\phi=-21.1^{\circ},62.6^{\circ}$~\cite{Zhao:2023hxc} \footnotemark[4]    \\
\bottomrule[1.00pt]
\bottomrule[1.00pt]
\end{tabular*}
\footnotetext[1]{Extracted from the ratio (1.04) of those two di-electronic widths.}
\footnotetext[2]{Fitting the di-electronic width of the $\psi(4160)$.}
\footnotetext[3]{Extracted from the ratio (1.79) of those two di-electronic widths.}
\footnotetext[4]{Fitting the di-electronic widths of the $\psi(4040)$ and $\psi(4160)$.} 
\end{table}
Although various methods are employed to determine the mixing angles, their results indicate that a significant $3S$-$2D$ mixing effect exists. 
Understanding the nature of the mixing angle is essential, as it has an influence on the assignment of higher vector charmonium states, their properties of mass, production, and decay. 
Therefore, given the significant $3S$-$2D$ mixing effect observed in the charmonium states $\psi(4040)$ and $\psi(4160)$, it becomes crucial to identify the underlying dynamical mechanisms responsible for generating such strong mixing.

In this work, we focus on this issue. In the traditional potential model, the tensor term in the effective potential contributes to $S$-$D$ mixing in charmonium, but its effect is too weak to produce a large mixing angle. This necessitates exploring alternative mechanisms.
With advances in understanding unquenched effects in hadron spectroscopy, higher-mass states can no longer be accurately described using the quenched approximation. Among these effects, the coupled-channel mechanism has emerged as particularly significant, signaling that hadron spectroscopy is entering a high-precision era \cite{Chen:2016qju,Chen:2016spr,Liu:2019zoy,Mai:2022eur,Liu:2024uxn}. Leveraging these developments, we investigate whether coupled-channel effects could explain the large $3S$-$2D$ mixing angle reported in earlier studies. However, we find a notable discrepancy between our calculations and prior results: Our analysis yields mixing angles of $\theta_1=7^\circ$ and $\theta_2=10^\circ$, differing substantially from previous values.

This discrepancy cannot be overlooked. To resolve it, we scrutinize earlier methodologies and identify a key sensitivity: The mixing angle depends critically on the measured dielectronic widths of the $\psi(4040)$ and $\psi(4160)$. Unfortunately, existing experimental data are insufficient for a precise extraction of the mixing angle. Our findings underscore the urgent need for higher-precision measurements to clarify the $S$-$D$ mixing problem in the $\psi(4040)$ and $\psi(4160)$.

The paper is organized as follows. In Sec. \ref{secII},
we calculate the mixing angles of relevant states by introducing a coupled-channel mechanism.
In Sec. \ref{secIII}, we analyze the origin of the discrepancy between our calculations and prior determinations of the mixing angle for the charmonium states $\psi(4040)$ and $\psi(4160)$. A summary is presented in Sec. \ref{secIV}.

\section{$3S$-$2D$ Mixing scheme for $\psi(3^3S_1)$ and $\psi(2^3D_1)$ }\label{secII}

As mentioned above, charmonia $\psi(3770)$ and $\psi(3686)$ can be well understood by incorporating $2S$-$1D$ mixing \cite{Rosner:2001nm}.  Similar to the case of $\psi(3686)$ and $\psi(3770)$, the relationship in a $3S$-$2D$ mixing scheme is expressed via the following mixing scheme
\begin{gather}\label{mix-matric}
	\begin{pmatrix}
		|\psi^{\prime}_{3S\text{-}2D}\rangle \\ \psi^{\prime\prime}_{3S\text{-}2D}\rangle
	\end{pmatrix}=
	\begin{pmatrix}
		\cos\theta & -\sin\theta \\ \sin\theta & \cos\theta
	\end{pmatrix}
	\begin{pmatrix}
		| \psi(3 ^3S_1) \rangle \\ | \psi(2 ^3D_1) \rangle
	\end{pmatrix},
\end{gather}
where $\theta$ is the mixing angle\footnote{
Two commonly used mixing conventions exist for the $\psi(3^3S_1)$ and $\psi(2^3D_1)$  charmonium states. In Refs. \cite{Chao:2007it,Li:2009zu,Zhao:2023hxc,Bokade:2024tge}, the physical states are expressed using a mixing angle $\phi$:
\begin{gather}
	\begin{pmatrix}
		|\psi^{\prime}_{3S\text{-}2D}\rangle \\ |\psi^{\prime\prime}_{3S\text{-}2D}\rangle
	\end{pmatrix}=
	\begin{pmatrix}
		\cos\phi & \sin\phi \\ -\sin\phi & \cos\phi
	\end{pmatrix}
	\begin{pmatrix}
		| \psi(3 ^3S_1) \rangle \\ | \psi(2 ^3D_1) \rangle
	\end{pmatrix}.
\end{gather}
In contrast, other studies (e.g., Refs. \cite{Badalian:2008dv,Wang:2022jxj}) employ the convention described in Eq. (\ref{mix-matric}). This work adopts the latter convention. Note that the mixing angle $-\phi$ from Refs. \cite{Chao:2007it,Li:2009zu,Zhao:2023hxc,Bokade:2024tge} is equivalent to our $\theta$ (i.e., $-\phi = \theta$).
}.
Our convention is consistent with Refs. \cite{Badalian:2008dv,Wang:2022jxj}.
The $\psi^{\prime}_{3S\text{-}2D}$ and $\psi^{\prime\prime}_{3S\text{-}2D}$ correspond to the charmonium  states $\psi(4040)$ and $\psi(4160)$, respectively.

The mixing dynamics between the $\psi(3^3S_1)$ and $\psi(2^3D_1)$ states can be induced by the tensor interaction within the potential model,
which originates from the one-gluon-exchange force \cite{DeRujula:1975qlm}. This interaction is described by 
 \begin{equation}
	H_T(r)=\frac{4\alpha_s}{3r_{ij}^3} S_{ij},
\end{equation}
where $S_{ij}$ is the tensor operator, which is defined as
\begin{equation}
	S_{ij}=\frac{3\mathbf{S}_i\cdot \mathbf{r}_{ij}\mathbf{S}_j\cdot\mathbf{r}_{ij}}{r^2_{ij}}-\mathbf{S}_i\cdot\mathbf{S}_j,
\end{equation}
with the $\mathbf{S}_i$ and $\mathbf{S}_j$ denoting the spin of the charm quark and anticharm quark, respectively.
When acting on the $|LSJ\rangle$ basis, the tensor operator $S_{ij}$ transforms the states as follows:
\begin{gather}
\begin{pmatrix}
\Phi_{LS}(3^3S_1)^\dagger &  \Phi_{LS}(2^3D_1)^\dagger
\end{pmatrix}
	S_{ij}
        \begin{pmatrix}
		\Phi_{LS}(3^3S_1) \\  \Phi_{LS}(2^3D_1) 
	\end{pmatrix}=
	\begin{pmatrix}
		0 & 2\sqrt{2}\\ 2\sqrt{2} & -2
	\end{pmatrix}.\nonumber\\
\end{gather}
Here, $\Phi_{LS}(3^3S_1)$ and  $\Phi_{LS}(2^3D_1)$ represent the $|LSJ\rangle$ basis for the $3S$-wave and $2D$-wave charmonium states, respectively.
To quantify the $3S$-$2D$ mixing angle, we employ the Godfrey-Isgur (GI) model, with details provided in Ref. \cite{Godfrey:1985xj}. The Hamiltonian is expressed as 
\begin{equation}
	H=T+\tilde{V}_{eff}(\mathbf{P},\mathbf{r}),
\end{equation}
where $T$ is the kinetic term, and the effective potential $\tilde{V}_{eff}(\mathbf{P},\mathbf{r})$  incorporates the tensor term $H_{T}(r)$.
 By solving the coupled-channel Schr$\ddot{\text{o}}$dinger function with this Hamiltonian, we can determine the $3S$-$2D$ mixing angle. 
Our calculations show that this approach typically predicts a mixing angle of only around $0.3^{\circ}$ between $\psi(3^3S_1)$ and $\psi(2^3D_1)$ states. This value is far too small to account for the experimentally observed properties, indicating that more realistic dynamics for higher charmonium states have not been adequately considered.

For higher excited charmonium states, coupled-channel effects present a more significant source of mixing, capable of generating a larger mixing angle. For instance, our prior work in Ref. \cite{Man:2025zfu} demonstrated a large mixing angle between $\psi(4220)$ and $\psi(4380)$, attributed to mixing dynamics induced by coupled-channel effects. %Additionally, coupled-channel effects significantly influence the masses of states lying above the open-charm threshold.

Borrowing this experience, we investigate this mixing scheme for the charmonium  states $\psi(3^3S_1)$ and $\psi(2^3D_1)$  within a coupled-channel framework, where the masses of the mixing states and the corresponding mixing angles are determined by solving the following equation:

\begin{eqnarray}\label{mix}
	\text{det}\begin{vmatrix}
		M_S^0 + \Delta M_{S}(M) - M &
        \langle \psi_S | H_{\text{T}} | \psi_D \rangle + \Delta M_{SD}(M) \\
		\langle \psi_D | H_{\text{T}} | \psi_S \rangle + \Delta M_{SD}(M)  &
        M_D^0 + \Delta M_{D}(M)- M
	\end{vmatrix} = 0.\nonumber\\
 \label{haha}
\end{eqnarray}
In this expression, $M^0_{S(D)}$ represents the bare masses for the $\psi(3^3S_1)$ and $\psi(2^3D_1)$ states, obtained from the GI model \cite{Godfrey:1985xj}. The terms $\Delta M_{S(D)}(M)$ denote the mass shifts for the $\psi(3^3S_1)$ and $\psi(2^3D_1)$ states induced by virtual meson loops, while the off-diagonal term $\Delta M_{SD}(M)$ provides the dominant contribution to the mixing between $\psi(3^3S_1)$ and $\psi(2^3D_1)$. The matrix element $\langle \psi_S | H_{\text{T}} | \psi_D \rangle$ represents the tensor term from the GI  model, which contributes only minimally to the $S$-$D$ mixing angle as mentioned above.

In principle, all possible virtual open-charm meson channels should be included in the calculation. However, accounting for an infinite number of virtual coupled channels in $\Delta M_{S(D)}(M)$ and $\Delta M_{SD}(M)$ presents a significant challenge. To address this, we introduce a once-subtracted dispersion relation to suppress contributions from distant virtual coupled channels. The expressions for $\Delta M_{S(D)}(M)$ and $\Delta M_{SD}(M)$ then become:
\begin{equation}\label{oncePiBC}
	\Delta M_{S(D)}(M)
	=\text{Re} \sum_{BC,JL}\int_0^\infty\frac{(M_{J/\psi}-M)|\mathcal{M}_{JL} ^{\psi_{S(D)}}|^2P^2 dP}
	{(M-E_B-E_C)(M_{J/\psi}-E_B-E_C)},
\end{equation} and
\begin{equation}
	\Delta M_{SD}(M)= \text{Re} \sum_{BC,JL}\int_0^\infty\frac{(M_{J/\psi}-M)\mathcal{M}_{JL} ^{\psi_{S}*}\mathcal{M}_{JL} ^{\psi_{D}}P^2dP}{(M-E_B-E_C)(M_{J/\psi}-E_B-E_C)},
\end{equation}
respectively.
In this work, we consider $DD$, $DD^{*}$, $D^{*}D^{*}$, $D_sD_s$, and $D_sD_s^*$ as the primary coupled channels for $\psi(3^3S_1)$ and $\psi(2^3D_1)$, as their respective thresholds lie below the bare mass of $\psi(3^3S_1)$. The transition amplitude $\mathcal{M}_{JL} ^{\psi_{S(D)}}=\langle\psi_{(S(D))}| H_I|BC\rangle$ describes the interaction between the bare state and the virtual hadronic channels. We adopt the quark pair-creation (QPC) model to calculate $\mathcal{M}_{JL}^{\psi_{S(D)}}$ \cite{LeYaouanc:1972vsx,LeYaouanc:1973ldf}. The subtraction point $M_{J/\psi}$ is taken as the mass of the $J/\psi$ meson in the charmonium system. Previous studies using this method have successfully reproduced reasonable meson masses, as demonstrated in Refs. \cite{Pennington:2007xr,Duan:2020tsx,Duan:2021alw,Deng:2023mza,Man:2025zfu,Zhou:2013ada}.

In order to obtain the physical mass, we should to solve the eigenvalue equation related to $M$ in Eq.~(\ref{mix}). For a general $n\times n$ Hermitian matrix, one could obtain $n$ eigenvalues and $n$ orthogonal eigenvectors. However, in Eq.~(\ref{mix}), the Hermitian matrix
\begin{equation}\label{eq:H}
H=\left(
\begin{array}{cc}
M_S^0 + \Delta M_{S}(M)&
\langle \psi_S | H_{\text{T}} | \psi_D \rangle + \Delta M_{SD}(M) \\
\langle \psi_D | H_{\text{T}} | \psi_S \rangle + \Delta M_{SD}(M)  &
M_D^0 + \Delta M_{D}(M)
\end{array}
\right)
\end{equation}
naturally contains eigenvalue $M$, which appears in $\Delta M_{S/D}(M)$. In this scheme, it is a nonlinear eigenvalue problem. For a given $M$, one could obtain two eigenvalues from Eq.~({\ref{eq:H}}), but the result which is  different from the given $M$ may  not satisfy the relationship in Eq.~({\ref{mix}}). In this scheme, with different physical solutions, the matrix in Eq.~(\ref{eq:H}) may have different matrix elements. Thus the physical eigenvectors may be not orthogonal, which reflect to the solutions with different mixing angles.

The spatial wave functions of the involved coupled channels play a crucial role in the numerical calculations. These wave functions are obtained by solving the Schrödinger equation with the GI model. For practical calculation, we represent the resulting spatial wave function in the following form:
\begin{eqnarray}\label{true wave function}
	\Psi_{ l m }(\mathbf{P}) &= \sum\limits_{n=1}^{n_{\text{max}}} C_{nl} \Phi_{nlm}(\mathbf{P}),
\end{eqnarray}
where $\Phi_{nlm}(\mathbf{P})$ represents a set of basis functions in momentum space, and $n_{\text{max}}=20$ is the maximum number of basis states used in the expansion.

Before implementing the $S$-$D$ mixing scheme, we quantitatively analyze the coupling strength of $\psi(3^3S_1)$ and $\psi(2^3D_1)$ to the virtual coupled channels.
For $\psi(3^3S_1)$ and $\psi(2^3D_1)$, the coupled-channel equation (\ref{mix}) can be decomposed as
\begin{equation}
M-M^0_{S(D)}-\Delta M_{S(D)}(M)=0.
\end{equation}

 When we use the parameter from Ref. \cite{Duan:2020tsx}, the bare masses of $\psi(3^3S_1)$ and $\psi(2^3D_1)$ are estimated to be 4089 and 4172 MeV, respectively.
According to our results in Table \ref{mass shift}, when the coupled-channel effects are taken into account , the bare mass of $\psi(3^3S_1)$ will shift by 81.8 MeV. The $\psi(3^3S_1)$ state predominantly couples to the $D^*D^*$ channel via the $P$ partial waves, while the $F$ waves provide negligible contributions.
The $DD$ ($D_sD_s$) and $DD^{*}$($D_sD_s^*$) channels provide relatively small contributions to the mass shift of $\psi(3^3S_1)$, mainly through the $P$ waves, respectively.

For the $\psi(2^3D_1)$ state, the physical mass we obtained (4087.3 MeV) is smaller than the PDG value \cite{ParticleDataGroup:2024cfk}.
However, a recent study by the Lanzhou group, based on an analysis of the $B^+\to K^+\mu^+\mu^-$ process, suggests that previous experiments overestimated the mass of the $\psi(4160)$ \cite{Peng:2024blp}. From Table \ref{mass shift}, our results show that the $\psi(2^3D_1)$ strongly couples with $DD$ and $D^*D^*$ channels via the $P$ wave. 
  
Furthermore, the  Okubo-Zweig-Iizuka (OZI)-allowed two-body strong decay width for the $\psi(3^3S_1)$ and $\psi(2^3D_1)$ can also be obtained by the amplitude $\mathcal{M}_{JL}^{S(D)}$, which is given by
\begin{eqnarray}
	\Gamma_\text{total}=\sum_{BC,JL} \frac{2\pi |\mathbf{P}|E_BE_C}{M}\mid\mathcal{M}_{JL}^{\psi_{S(D)}})\mid^2,
\end{eqnarray}
where $\mathbf{P}$ is the momentum of the final-state mesons and $M$ is the physical mass.

 From Table \ref{mass shift}, the open-charm decay width of the $\psi(3^3S_1)$ is predicted to be 45.1 MeV, with the main decay channel being $DD^*$.
 The predicted width of $\psi(3^3S_1)$ is 36.4 MeV, and it dominantly decays into $DD$, $DD^*$, and $D^*D^*$ channels.

\begin{table*}[!htbp]
\caption{The mass shifts ($\Delta M_i$) and decay widths ($\Gamma_i$) of the $\psi(3^3S_1)$ and $\psi(2^3D_1)$ states. The values are given in units of MeV.In this work, $DD$ is used as a shorthand notation to represent the sum of the $D\bar{D}$ and $\bar{D}D$ decay channels, and similar notations are adopted for other decay modes.}\label{mass shift}
\centering
\renewcommand\arraystretch{1.35}
\begin{tabular*}{2.0\columnwidth}{@{\extracolsep{\fill}}cccccccccc@{}}
\toprule[1.00pt]
\toprule[1.00pt]
Channel & $\Delta M_i$ & $\Delta M_i/\sum_i\Delta M_i$ & $\Gamma_i$ & $\Gamma_i/\sum_i\Gamma_i$ 
        & $\Delta M_i$ & $\Delta M_i/\sum_i\Delta M_i$& $\Gamma_i$ & $\Gamma_i/\sum_i\Gamma_i$ \\
\midrule[0.75pt]
$DD$          & $-6.2$  & 7.5\%  & 0.2  & 0.4\%  & $-16.4$ & 19.3\% & 7.6  & 21.0\% \\
$DD^*$        & $-$7.8  & 9.5\%  & 44.1 & 97.8\% & $-$1.2  & 1.4\%  & 6.2  & 17.1\% \\
$D_sD_s$      & $-$1.2  & 1.5\%  & 0.8  & 1.8\%  & $-$1.0  & 1.2\%  & 1.2  & 3.3\%  \\
$D^*D^*$      & $-$62.1 & 76.0\% & $\dots$    & $\dots$      & $-$64.0 & 75.6\% & 21.2 & 58.2\% \\
$D_sD_s^*$    & $-$4.5  & 5.5\%  & $\dots$    & $\dots$      & $-$2.1  & 2.4\%  & 0.2  & 0.5\%  \\
Total         & $-$81.8 & 100\%  & 45.1 & 100\%  & $-$84.7 & 100\%  & 36.4 & 100\%  \\
\bottomrule[1.00pt]
\bottomrule[1.00pt]
\end{tabular*}
\end{table*}

\begin{table}[!htbp]
\caption{The $3S$-$2D$ mixing angles for $\psi^{\prime}_{3S\text{-}2D}$ and $\psi^{\prime\prime}_{3S\text{-}2D}$ are induced by the coupled-channel effects. The mixing states are expressed as $C_{S1(2)}|\psi(3 ^3S_1)\rangle$ +$C_{D1(2)}|\psi(2 ^3D_1)\rangle$. }\centering
\renewcommand\arraystretch{1.35}\label{mixing states}
\begin{tabular*}{1.0\columnwidth}{@{\extracolsep{\fill}}cccc}
\toprule[1.00pt]
\toprule[1.00pt]
States                         & $C_{S1(2)}$  &$C_{D1(2)}$  &Mixing angle  \\
\midrule[0.75pt]
$\psi^{\prime}_{3S\text{-}2D}$       & $-0.992$ & $0.128$ &$\theta_1=7^{\circ}$  \\ 
$\psi^{\prime\prime}_{3S\text{-}2D}$ & 0.180  & 0.984 & $\theta_2=10^{\circ}$ \\
\bottomrule[1.00pt]
\bottomrule[1.00pt]
\end{tabular*}
\end{table}

From Eq.~(\ref{mix}), one can see that the $\Delta M_{SD}(M)$ term plays a dominant role in determining the mixing angle. The associated uncertainty is mainly governed by the quark pair-creation strength $\gamma$, which simultaneously affects the physical masses, the OZI-allowed two-body strong decay widths, and the mixing angles. To obtain reliable estimations, $\gamma$ is typically taken to be around $0.4$ \cite{Duan:2020tsx,Deng:2023mza}, as fitted from the decay widths of various charmonium states. If $\gamma$ deviates significantly from this value, the resulting physical masses and decay widths become unreasonable. To examine its impact, we consider three representative values $\gamma=0.37$, $0.40$, and $0.43$, and analyze their corresponding effects on $\psi^{\prime}_{3S\text{-}2D}$ and $\psi^{\prime\prime}_{3S\text{-}2D}$.

By solving Eq. (\ref{mix}) within our established framework, we obtain the physical masses of the mixing states $\psi^{\prime}_{3S\text{-}2D}$ and $\psi^{\prime\prime}_{3S\text{-}2D}$, which are $4006.4^{+7.4}_{-8.3} $ and $4089.5^{+8.8}_{-8.6}$ MeV, respectively. Additionally,
in Table \ref{mixing states}, we present the corresponding eigenvectors for these states. Based on the obtained coefficients $C_{S1(2)}$ and $C_{D1(2)}$, the mixing angles are extracted using the relations $\theta_1=\arctan(-C_{D1}/C_{S1})$ and $\theta_2=\arctan(C_{S2}/C_{D2})$ for $\psi^{\prime}_{3S\text{-}2D}$ and $\psi^{\prime\prime}_{3S-2D}$, respectively \cite{Ni:2025gvx}.
The eigenvectors $(C_{S1(2)}, C_{D1(2)})$ represent the mixing coefficients for the $3^3S_1$ and $2 ^3D_1$ components in each physical state.
Then, the mixing angles are determined to be
\begin{equation}
\theta_1=(7.3^{+0.1}_{-0.4})^{\circ}\quad{\text{for}}\quad \psi^{\prime}_{3S\text{-}2D},
\end{equation}
and 
\begin{equation} 
\theta_2=(10.4^{+2.0}_{-3.5})^{\circ} \quad{\text{for}}\quad \psi^{\prime\prime}_{3S\text{-}2D}.
\end{equation}

Based on the above discussion, the mixing angles remain relatively stable with the change of $\gamma$ and are determined to be around $7^{\circ}$ and $10^{\circ}$. For convenience, these two values are adopted in the subsequent numerical calculations.

\section{An Analysis of Potential Reasons for the Discrepancy}\label{secIII}

Our obtained results are notably smaller than those reported in Table \ref{mix-angle} by other theoretical studies \cite{Chao:2007it,Li:2009zu,Badalian:2008dv,Wang:2022jxj,Bokade:2024tge,Zhao:2023hxc}.
As shown in Table \ref{mix-angle}, we can find that derivation of most larger mixing angles relies on fitting the experimentally measured dielectronic widths of the $\psi(4040)$ and $\psi(4160)$ or their ratio.   
Consequently, the inferred mixing angle is highly dependent on the measured dielectronic widths of the $\psi(4040)$ and $\psi(4160)$. However, when we check the PDG \cite{ParticleDataGroup:2024cfk}, one finds that the reported values of the dielectronic decay width for the $\psi(4160)$ vary significantly across different experimental collaborations.
 Before 2008, the experimentally measured dielectronic decay width was approximately 0.8 keV \cite{Seth:2004py,DASP:1978dns}. Based on the ratio (1.04) of the dielectronic widths of the $\psi(4040)$ (0.86 keV) and $\psi(4160)$ (0.83 keV), theoretical predictions yielded a mixing angle of about $35^\circ$ \cite{Chao:2007it, Badalian:2008dv,Li:2009zu}. 
 
 In 2008, by considering the interference effect in the analysis of $R$ values, the BES Collaboration reported a higher mass and a smaller di-electronic width for the $\psi(4160)$, which are different from previous results \cite{BES:2007zwq}.
In 2010,  Mo, Yuan, and Wang reanalyzed BES data highlighted that the Ref. \cite{BES:2007zwq} reported dielectronic width of the $\psi(4160)$ is just one of four possible solutions,  with the values ranging from 0.4 to 1.1 keV \cite{Mo:2010bw}.  It is crucial to note that the existence of multiple possible solutions for the dielectronic width introduces significant uncertainties into the extraction of the $3S$-$2D$ mixing angle. 
In Refs. \cite{Wang:2022jxj,Zhao:2023hxc} , the authors estimated the $3S$-$2D$ mixing angle to  be around $20^{\circ}$ by using a screening potential model and the measured di-electronic width of $0.48\pm 0.22$ keV.
 A significant $3S$-$2D$ mixing angle was also obtained  \cite{Bokade:2024tge}, derived from the ratio (1.79) of the dielectronic widths of the $\psi(4040)$ (0.86 keV) and $\psi(4160)$ (0.48 keV).

Historically, early measurements of the $\psi(4040)$ and $\psi(4160)$ were  primarily based on inclusive processes in $e^+e^-\to \text{hadrons}$, with limited contributions from exclusive decay channels. Especially, the  dielectronic decay width of the $\psi(4160)$ extracted from inclusive data exhibits significant uncertainties. This is largely because the numerous open-charm thresholds in this energy region lead to complex structures in the cross-section behavior. The inclusive measurements do not provide sufficient experimental information for a detailed understanding  the structure of the high-mass charmonium states. 
Furthermore, although two exclusive processes involving the $\psi(4160)$ \cite{LHCb:2013ywr,BESIII:2023wsc}, its dielectronic decay width has yet to be precisely determined. Consequently, methods that depend on measured dielectronic widths for determining the mixing angles between $\psi(3^3S_1)$ and $\psi(2^3D_1)$ may lead to unreliable conclusions.
Facing these challenges, a more comprehensive understanding of the internal structure and mixing dynamics of the $\psi(4040)$ and $\psi(4160)$ is essential.

To better understand the mixing states $\psi^{\prime}_{3S\text{-}2D}$ and $\psi^{\prime\prime}_{3S\text{-}2D}$, we investigate their OZI-allowed two-body strong decay widths and dielectronic widths, utilizing our calculated mixing angles.
The OZI-allowed two-body strong decay width for the $\psi_{3S-2D}^{\prime}$ and $\psi_{3S-2D}^{\prime\prime}$ take the form
\begin{equation}
	\Gamma^{\psi^{\prime}_{3S\text{-}2D}}_\text{total}=\sum_{BC,JL} \frac{2\pi |\mathbf{P}|E_BE_C}{M}\left|(\cos\theta_1\mathcal{M}_{JL}^{\psi_{S}}-\sin\theta_1\mathcal{M}_{JL}^{\psi_{D}})\right|^2,
\end{equation}
\begin{equation}
	\Gamma^{\psi^{\prime\prime}_{3S\text{-}2D}}_\text{total}=\sum_{BC,JL} \frac{2\pi |\mathbf{P}|E_BE_C}{M}\left|(\sin\theta_2\mathcal{M}_{JL}^{\psi_{S}}+\cos\theta_2\mathcal{M}_{JL}^{\psi_{D}})\right|^2.
\end{equation}
From Table \ref{mix-decay}, based on the mixing angle, the two-body strong decay decay width  of $\psi^{\prime}_{3S\text{-}2D}$ is estimated to be $38.0$ MeV. This value is smaller than the PDG value \cite{ParticleDataGroup:2024cfk}. However, 
it is crucial to note that the PDG width for the $\psi(4040)$ is predominantly based on inclusive measurements, rather than exclusive decay channels.  Therefore, a direct comparison between our calculated two-body decay width and the PDG's total  width is not straightforward, as the latter encompasses contributions from numerous unmeasured or complex decay modes.

The two-body strong decay width of $\psi^{\prime\prime}_{3S\text{-}2D}$  is predicted to be $31.1$ MeV, which is consistent with the BESIII exclusive measured result $55\pm15\pm53$ MeV \cite{BESIII:2023wsc}. Because of the relatively small mixing angle obtained in our model, the dominant decay channel for these states is predicted to remain largely unchanged from that of their unmixed states, supporting the standpoint that their primary two-body strong decay properties are preserved despite mixing.

\begin{table}[!htbp]
\renewcommand\arraystretch{1.5}
	\caption{Two-body strong decay widths of ${\psi}_{3S-2D}^\prime$ and ${\psi}_{3S-2D}^{\prime\prime}$ with mixing angles $7^{\circ}$ and $10^{\circ}$, respectively. The decay widths are in units of MeV. }\centering
 \renewcommand\arraystretch{1.35}
	\begin{tabular*}{1.0\columnwidth}{@{\extracolsep{\fill}}ccccc@{}}
    \toprule[1.00pt]
    \toprule[1.00pt]
    \label{mix-decay}
		&  \multicolumn{2}{c}{${\psi}_{3S\text{-}2D}^\prime$}  &\multicolumn{2}{c}{${\psi}_{3S\text{-}2D}^{\prime\prime}$}\\  
  	& \multicolumn{2}{c}{ $\theta_1=7^\circ$}  & \multicolumn{2}{c}{$\theta_2=10^\circ$}\\
   \midrule[0.75pt]
		Channels &$\Gamma_{i}$&$\Gamma_{i}/\sum_i \Gamma_{i}$&$\Gamma_{{i}}$&$\Gamma_{i}/\sum_i \Gamma_{i}$\\ \hline 
$DD$          & 0.2  & 0.5\%  & 5.7  & 18.2\%\\
$DD^*$        & 36.7 & 96.7\% & 8.3  & 26.6\%\\
$D_sD_s$      & 1.1  & 2.8\%  & 0.9  & 2.8\% \\
$D^*D^*$      & $\dots$  & $\dots$   & 15.9 & 51.1\%\\
$D_sD_s^*$    & $\dots$  & $\dots$   & 0.4  & 1.3\% \\
Total      & 38.0   & 100\%  & 31.1      & 100\% \\
       \bottomrule[1.00pt]
       \bottomrule[1.00pt]
	\end{tabular*}
\end{table}

The dielectronic widths of the mixed $\psi^{\prime}_{3S\text{-}2D}$ and $\psi^{\prime\prime}_{3S\text{-}2D}$, including radiative QCD corrections \cite {Kwong:1987ak,Wang:2019mhs},  are given by
\begin{equation}\label{eq:Gamma1}
	\Gamma^{\psi^{\prime}_{3S\text{-}2D}}_{e^+e^-}=\frac{16\pi\alpha^2 e_c^2 \mathcal{C}}{M^2_{\psi^{\prime}_{3S\text{-}2D}}}\left|\text{cos} \theta_1 R_{3S}(0)-\frac{5\sqrt{2}}{M^2_{\psi^{\prime}_{3S\text{-}2D}} }  \text{sin}\theta_1 R^{\prime\prime}_{2D}(0)\right|^2,
\end{equation}
\begin{equation}\label{eq:Gamma2}
	\Gamma^{\psi^{\prime\prime}_{3S\text{-}2D}}_{e^+e^-}=\frac{16\pi\alpha^2 e_c^2 \mathcal{C}}{M^2_{\psi^{\prime\prime}_{3S\text{-}2D}}}\left| \text{sin}  \theta_2 R_{3S}(0)+\frac{5\sqrt{2}}{M^2_{\psi^{\prime\prime}_{3S\text{-}2D}} } \text{cos}\theta_2 R^{\prime\prime}_{2D}(0)\right|^2,
\end{equation}
where $\alpha=1/137$ is the fine-structure constant, and $e_c=2/3$ denotes the charm quark charge in units of the electron charge, respectively. The factor $\mathcal{C}=(1-16\alpha_s/3\pi)$ represents the first-order QCD radiative correction. $R_{3S}(0)=0.16$ GeV$^{3/2}$ accounts for the radial $S$-wave function at the origin,  and $R_{2D}^{\prime\prime}(0)=0.11$ GeV$^{7/2}$ is the second derivative of the radial $D$-wave function at the origin.

\begin{figure}[!htbp]
	\centering
	\begin{minipage}[b]{0.48\textwidth}
		\centering
		\includegraphics[width=1\textwidth]{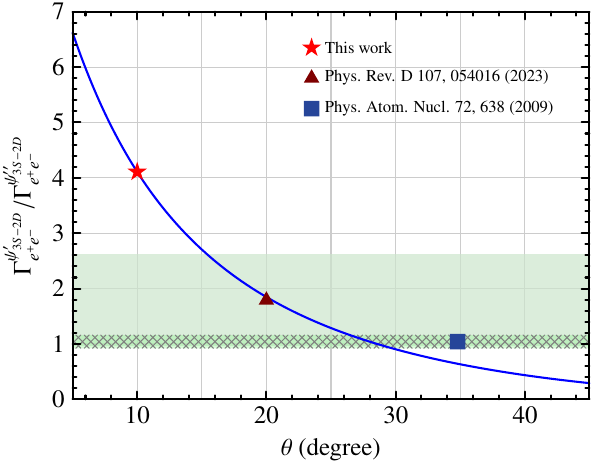}
	\end{minipage}
	\caption{The ratio of dielectronic widths $\Gamma^{\psi^{\prime}_{3S\text{-}2D}}_{e^+e^-}/\Gamma^{\psi^{\prime\prime}_{3S\text{-}2D}}_{e^+e^-}$ as a function of the $3S\text{-}2D$ mixing angles with our parameters from Eqs.~(\ref{eq:Gamma1})and (\ref{eq:Gamma2}). The red pentagram represents the prediction of this work ($\theta \approx 10^\circ$ and ratio $\Gamma^{\psi^{\prime}_{3S\text{-}2D}}_{e^+e^-}/\Gamma^{\psi^{\prime\prime}_{3S\text{-}2D}}_{e^+e^-}\approx 4.1$). The  light green and dark green shaded bands indicate the experimental values $1.79 \pm 0.83$ and $1.04 \pm 0.12$, respectively, from the PDG \cite {ParticleDataGroup:2024cfk}. 
 The triangle and square indicate the mixing angles $20^{\circ}$ and $34.8^{\circ}$, as obtained in Refs. \cite{Wang:2022jxj} and \cite{Badalian:2008dv}, respectively.
}  \label{ratio1}
\end{figure}

Using a mixing angle of $\theta_1 = 7^\circ$, the dielectronic decay width of $\psi^{\prime}_{3S\text{-}2D}$ is calculated to be $0.92~\text{keV}$. This result shows  agreement with both the PDG value of $0.86 \pm 0.07~\text{keV}$~\cite{ParticleDataGroup:2024cfk} and the experimentally measured range of $0.6$-$1.4~\text{keV}$ reported in Ref.~\cite{Mo:2010bw}.

For $\theta_2 = 10^\circ$, we predict a dielectronic width of $0.22~\text{keV}$ for $\psi^{\prime\prime}_{3S\text{-}2D}$, which approaches the lower experimental bound for the $\psi(4160)$ dielectronic decay width.

As shown in Fig.~\ref{ratio1}, our theoretical calculation gives a ratio of $\sim\!4.1$, significantly higher than the experimental values of $1.04 \pm 0.12$ and $1.79 \pm 0.83$~\cite{ParticleDataGroup:2024cfk}. This discrepancy highlights an important testable prediction for future high-precision experiments.

\section{Summary}\label{secIV}

Despite five decades of research since the $J/\psi$ discovery, the properties of charmonium states above the $D\bar{D}$ threshold, including the $\psi(4040)$ and $\psi(4160)$, remain enigmatic due to experimental and theoretical limitations. While these states are well documented in the PDG, their underlying physics is not fully understood. An intriguing feature is $S$-$D$ mixing, a phenomenon observed across charmonium states, with former studies proposing $3S$-$2D$ mixing for the $\psi(4040)$ and $\psi(4160)$ \cite{Chao:2007it,Li:2009zu, Badalian:2008dv,Anwar:2016mxo,Yang:2018mkn,Wang:2022jxj,Bokade:2024tge,Zhao:2023hxc}. Current determinations of the mixing angle \cite{Chao:2007it,Li:2009zu, Badalian:2008dv,Wang:2022jxj,Bokade:2024tge,Zhao:2023hxc}, derived from dielectronic widths or their ratios at different experimental periods \cite{Seth:2004py,BES:2007zwq}, vary significantly, indicating substantial mixing effects.

This work investigates the dynamical origins of this mixing. Traditional potential models, relying on tensor forces, cannot account for the large observed angles, prompting exploration of alternative mechanisms. Incorporating unquenched effects, particularly coupled-channel dynamics, we calculate mixing angles of $\theta_1=7^\circ$ and $\theta_2=10^\circ$ for $\psi^{\prime}_{3S\text{-}2D}$ and $\psi^{\prime\prime}_{3S\text{-}2D}$, respectively-markedly smaller than prior estimates. The discrepancy highlights a critical dependence on experimental dielectronic widths, where current data are inadequate for precise angle extraction. Our results emphasize the necessity of high-precision measurements to resolve the $S$-$D$ mixing puzzle in $\psi(4040)$ and $\psi(4160)$, advancing the broader understanding of charmonium spectroscopy.

The current upgrade of BESIII experiment is expected to achieve a threefold increase in luminosity \cite{Balossino:2022ywn}. This advancement will enable precision measurements of the spectroscopic properties of the $\psi(4040)$ and $\psi(4160)$, providing crucial experimental tests for the theoretical predictions presented in this work.
 
\begin{acknowledgments}
X.L. thanks Cheng-Ping Shen 
for his questions on my plenary talk at the 10th $XYZ$ Symposium held in Changsha, China. This study addresses his questions.
This work is supported by the National Natural Science Foundation of China under Grant No. 12335001, No. 12247101, and No. 12405098,  the "111 Center" under Grant No. B20063, the Natural Science Foundation of Gansu Province (No. 22JR5RA389 and No. 25JRRA799), the Talent Scientific Fund of Lanzhou University, the fundamental Research Funds for the Central Universities (No. lzujbky-2023-stlt01), the project for top-notch innovative talents of Gansu province, and Lanzhou City High-Level Talent Funding.
\end{acknowledgments}

\bibliography{inport}

\end{document}